\documentstyle[12pt]{article}        
\def\o{\over}   
\def\Ar{\rightarrow}        
\def\bar{\overline}

\def\d{\delta}        
\def\a{\alpha}        
\def\b{\beta}        
\def\n{\nu}        
\def\m{\mu}        
        
\def\e{\epsilon}        
        
\def\th{\theta}

\def\bar{\overline}        
\def\l{\lambda}

\def\eV{{\rm eV}}        
\begin{document}        
\baselineskip=24.5pt        
\setcounter{page}{1}        
\thispagestyle{empty}        
\topskip 1  cm  
\setcounter{page}{1}        
\thispagestyle{empty}        
\topskip 0.  cm        
\begin{flushright}        
Lund-Mph-97/14\\    
November 1997        
\end{flushright}        
\vspace{0.3 cm}        
\centerline{\Large\bf  Natural Neutrino Mass Matrix}        
\vskip 1.2 cm        
\begin{center}        
{\bf  Masahisa MATSUDA        
\footnote{Permanent address:Department of Physics and Astronomy,     
 Aichi University of Education, Kariya,        
\par         
Aichi 448, Japan. \ \         
E-mail address:mmatsuda@matfys.lth.se}${}^a$}        
{\bf  and}        
{\bf Morimitsu TANIMOTO        
\footnote{E-mail address: tanimoto@edserv.ed.ehime-u.ac.jp}${}^b$}        
\end{center}        
\vskip 0.5 cm        
\centerline{${}^a$ \it{Department of Mathematical Physics, LTH,     
Lund University, S-22100, Lund, Sweden}}        
\centerline{${}^b$ \it{Science Education Laboratory, Ehime University,         
 790-77 Matsuyama, JAPAN}}        
\vskip 2 cm        
\centerline{\bf ABSTRACT}\par     
\vskip 0.5 cm       
Naturalness of  the neutrino mass hierarchy and mixing        
 is studied.        
First we select among 12 neutrino mixing patterns a few patterns, which        
could form  the natural neutrino mass matrix.         
Further we show that if the Dirac neutrino mass matrix is taken    
as the natural one in the quark sector, then only two mixing patterns  
without the large mixing lead        
to the natural right-handed Majorana mass matrix. The rest of
 the chosen 
 patterns      with three degenerate mass solution    
lead to the unnatural right-handed        
Majorana mass matrix  in the see-saw mechanism.     
Notice however, that for the chosen  two natural patterns
there could  be a huge mass hierarchy such as 
${\cal O}(10^{4\sim 6})$         
 in order to reproduce the inverse mass hierarchy of the light neutrinos.        
\vskip 0.5 cm        
\newpage        
\topskip 0.  cm        
\voffset = -1 cm        
\hoffset = -.50 in        
\section{Introduction}        
\vskip 0.3 cm        
The standard model (SM) being extensively treated still 
has many unexplained features.
It contains 18 parameters, among which 13 parameters  belong to 
the flavor sector of the fermion masses and mixing \cite{KM}.        
It is a remarkable property of the quark and the lepton mass spectra that 
the masses of successive particles  increase by large factors.     
Mixings of the quark sector seem also to have an hierarchical structure.      
Those features may provide an important basis for a new physics beyond the SM.        
\par          
It is known that there are 54 parameters in the quark-lepton mass     
matrices:$M_U$, $M_D$ and $M_E$.        
Since only 13 combinations of 54 parameters are observable,        
the flavor sector is a highly undetermined system.        
In order to understand the pattern of fermion masses,         
it is necessary to reduce the number of arbitrary parameters        
in the Yukawa matrices from 54 to a number which is less than 13.        
Therefore, one  asks for  an organizing principle for the mass matrices.        
The idea of {\it natural mass matrix}, elaborated by Peccei 
and Wang \cite{Natu}, severely restricts the     
arbitrariness in the construction of the quark mass matrices.        
Here we study the natural mass matrix in the lepton sector using essentially
the recent neutrino oscillation experiments \cite{solar}$\sim$\cite{CHOOZNew}.
\par         
Neutrino flavor oscillations are important phenomena for the study of the 
neutrino masses and mixing.        
The only possible evidences for neutrino oscillations originate from the     
natural beams:        
the solar neutrinos \cite{solar} and the atmospheric 
neutrinos \cite{Atm1}$\sim$\cite{SKam}.        
Data from the accelerator and the reactor neutrino experiments will be 
available 
in the near future.     
In fact data  given by the LSND experiment \cite{LSND} indicate     
at least one large mass difference.
More precisely, one can expect that $\Delta m^2 \sim {\cal O}(1~\eV^2)$ 
with $\sin^2 2\th_{\rm LSND}\simeq 10^{-3}$.
We expect data from the KARMEN experiment \cite{KARMEN} where the oscillations 
$\n_\m\Ar\n_e(\bar \n_\m\Ar \bar \n_e)$ are being considered as in LSND.
Also we expect data  from the  CHORUS and  NOMAD experiments \cite{CHONOM} 
on the $\n_\m\Ar\n_\tau$ oscillation.         
The  most powerful reactor experiments searching for the neutrino     
oscillation are  those of Bugey \cite{Bugey}  and 
Krasnoyarsk \cite{Krasnoyarsk} at present.     
These results show the region in the  parameter space $(\sin^2 2\th, \Delta m^2)$ 
where the neutrino oscillation is not observed.        
The first long baseline reactor experiment CHOOZ \cite{CHOOZ,CHOOZNew} also gives 
us severe constraints.        
The possibilities of the accelerator long baseline (LBL) experiments     
have been discussed for example, in \cite{long,KEKPS}.        
The first experiment will begin in KEK-SuperKamiokande (K2K) \cite{KEKPS}.    
Presently, the constraints are also given by the disappearance experiments    
in CDHS \cite{CDHS} and CCFR \cite{CCFR},  and the appearance experiments 
E531 \cite{E531}, E776 \cite{E776}  and new CCFR \cite{CCFRNEW}.        
The bounds for the neutrino masses and mixing are given 
in \cite{Mina}$\sim$\cite{tani} on the basis of the listed experiments.     
\par        
Still, the neutrino masses and mixing are not determined uniquely.        
There should be only two hierarchical mass difference scales $\Delta m^2$      
in the three-flavor mixing scheme.        
If the highest neutrino mass scale is chosen to be the  LSND scale 
${\cal O}(1\eV)$, the other mass scale should be either the atmospheric neutrino 
mass scale $\Delta m^2\simeq 10^{-2}\ \eV^2$ \cite{Atm1}$\sim$\cite{SKam} or one 
of the solar neutrino mass scale: $\Delta m^2\simeq 10^{-5}\sim 10^{-6}\ \eV^2$
(MSW solution) or $10^{-10}\sim 10^{-11}\ \eV^2$(just-so solution) \cite{solar}.
In this case, the expected neutrino masses and mixing depend on the experimental 
data.
Since the neutrino masses and mixing angles are determined independently      
in those experiments, the principle of the naturalness can help us to select 
a few patterns of the masses and mixings among  many patterns which are available 
from the present experimental data.        
Thus, we can suggest the experimental data which conduce to     
the natural neutrino mass matrix.        
\par    
Our paper is organized as follows.  
In section 2  the concepts of the natural mass matrix are discussed following 
the arguments provided in Ref.\cite{Natu} for small angle mixing case. 
Further we generalize these concepts to the large angle mixing case.    
In section 3  we study possible patterns of the neutrino masses and mixing      suggested by the recent neutrino experiments.
In section 4  the natural left-handed Majorana mass matrices are     
investigated using the  patterns obtained in section 3.
In section 5  a  natural see-saw realization  is considered based on the 
right-handed Majorana neutrinos.        
Finally, section 6 is devoted to the conclusions.        
\vskip 0.5 cm          
\section{Naturalness of the mass matrix}    
\vskip 0.3 cm     
The idea of the natural mass matrix was proposed in order to restrict     
severely the arbitrariness in the construction of the quark 
mass matrices \cite{Natu}.    
This idea was worked out in the context of SUSY GUT's, when some potentially 
successful  quark mass matrices were found at the GUT scale.        
\par    
Let us consider the $2\times 2$ Hermitian quark mass matrix $M_i(i=u,\ d)$.     
Assume for simplicity that it can be diagonalized by some orthogonal matrices 
$O_i(i=u,\ d)$  as follows:     
\begin{equation}          
   O_i^T M_i O_i= M_i^{diag}\equiv \left (\matrix{ m_{i1} & 0 \cr      
                 0 & m_{i2} \cr   }\right ) \ , \qquad  (i=u,\ d ) \ ,      
  \end{equation}          
\noindent     
with         
\begin{equation}         
   O_i=\left (\matrix{    
                \cos \th_i & \sin\th_i \cr     
               -\sin \th_i & \cos \th_i \cr    
                      } \right ), \qquad           
       V_q=O_u^T O_d=\left (\matrix{    
                \cos \th_c & \sin\th_c \cr     
               -\sin \th_c & \cos \th_c \cr    
                      } \right ),        
\end{equation}        
\noindent        
where $\th_c=\th_d-\th_u$.     
Then the  mass matrices can be written as    
\begin{equation}          
    M_i=  \left (\matrix{ c_i^2 m_{i1}+  s_i^2 m_{i2} &  c_i s_i(m_{i2}- m_{i1}) \cr      
                 c_i s_i(m_{i2}- m_{i1}) & s_i^2 m_{i1}+  c_i^2 m_{i2} \cr   }\right ) \ ,     
				 \qquad  (i=u,\ d ) \ ,    
	\label{Matrix}      
  \end{equation}      
 \noindent    
 where $c_i\equiv \cos \th_i$ and $s_i\equiv\sin \th_i$.    
   The mixing matrix $V_q$ is invariant under the changes        
      \begin{equation}         
  O_d\Ar O O_d \ , \qquad\qquad    O_u\Ar O O_u \ ,      
  \label{Change}       
\end{equation}        
\noindent     
where $O$ is some arbitrary orthogonal matrix.        
Thus, using the fact that $\sin\th_c\ll 1$, we can assume both 
$\th_d\ll 1$ and $\th_u\ll 1$ without loss of generality.     
Taking into account the quark mass hierarchy, we set    
\begin{equation}          
  m_u= a\l^4  m_c \ , \qquad   m_d= b\l^2  m_s \ ,    
  \end{equation}     
  \noindent    
   where $a$ and $b$ are ${\cal O}(1)$ coefficients, and    
$\l\equiv \sin\th_c  \simeq 0.22$. 
Thus the mass matrices are expressed in terms of $\th_u$ and $\th_d$ as following:
\begin{equation}         
 M_u\simeq \left (\matrix{    
              a\l^4+\sin^2\th_u & \sin\th_u \cr     
              \sin \th_u & 1\cr    
                   } \right )m_c \ , \qquad          
 M_d\simeq \left (\matrix{    
              b\l^2+\sin^2\th_d & \sin\th_d \cr     
              \sin \th_d & 1\cr    
                   } \right )m_s \ .        
\end{equation}        
There are three different options now for the angles $\th_u$ and $\th_d$:       
\noindent    
$$1.\ \sin\th_d\sim \l,\  \sin\th_u\sim \l, \quad        
  2.\ \sin\th_d\sim \l,\  \sin\th_u\leq \l^2, \quad       
  3.\ \sin\th_d\leq \l^2,\  \sin\th_u\sim \l.$$       
     
\noindent    
The options 1 and 3 are unnatural. 
Indeed, one would require a severe     
fine-tuning of the matrix element ${[M_u]}_{11}$ forcing     
${[M_u]}_{11}\simeq {([M_u]}_{12})^2$        
to arrive at the large $m_u/m_c\sim \l^4$ hierarchy.        
On the other hand, option 2 gives a natural quark mass matrix without fine-tuning.
Here the fine-tuning means a tuning of ${\cal O}(\l^2)$, which comes from the 
following  inspection.  
There is a  well known  phenomenological relation between the CKM  
matrix element and the quark mass ratio as  
\begin{equation}  
  |V_{us}|\simeq \sqrt{m_d\o m_s}\simeq \l \ .  
\end{equation}  
  \noindent  
The down quark mass ratio dominates  $V_{us}$, while the contribution of the up 
quark mass ratio is at most $\sqrt{m_u/ m_c}\simeq {\cal O}(\l^2)$.  
This  relation is consistent with the option 2.  
Thus, the criterion of ${\cal O}(\l^2)$  tuning is very useful in order to select 
the option 2 by excluding  options 1 and 3.  
The success of  Wolfenstein parameterization of the CKM matrix \cite{Wol} 
guarantees that this criterion is also available for the three generation case.  
The extension of the natural mass matrices to the three family case   
in quark sector is also given  in Ref.\cite{Natu}.    
\par    
Naturalness is also expanded to the lepton sector.   
If neutrino sector has the hierarchical mass structure, the naturalness argument 
is exactly the same as in the example just described for the quark sector.  
However,  in the neutrino sector  the inverse mass hierarchy for the flavor 
is still allowed by the constraints obtained in the disappearance experiments of 
the neutrino oscillations in Bugey \cite{Bugey}, Krasnoyarsk \cite{Krasnoyarsk}, 
CDHS \cite{CDHS} and CCFR \cite{CCFR}.     
Since the neutrino mixing is chosen to be $\sin\th_\n\sim 1$  in this case, one 
cannot always guarantee to have both $\th_\n\ll 1$ and $\th_E\ll 1$(charged 
lepton mixing) by the change in eq.(\ref{Change}).     
Therefore, we should reconsider  the naturalness of the quark mass  
matrices in the lepton sector.   
\par  
In the case of $s\simeq 1(c\ll 1)$ with $m_{1}\leq m_{2}$ in eq.(\ref{Matrix}), 
the mass matrix is expressed approximately:    
\begin{equation}          
    M_\nu\simeq  \left (\matrix{m_2 &  c m_2 \cr      
                 c m_2 & m_1+  c^2 m_2 \cr   }\right ) \ .     
\end{equation}     
The natural mass matrix without the fine-tuning requires     
\begin{equation}    
   m_1 \geq c^2 m_2 \ .    
\label{cond1}  
\end{equation}    
\noindent   
For example, if we take $c\simeq a'\l, m_1\simeq a\l m_2$, which satisfies the 
condition(\ref{cond1}), then matrix  
\begin{equation}         
 M_\nu\simeq \left (\matrix{    
               1 & a'\l \cr     
              a'\l & a\l \cr    
                   } \right )m_2  
\end{equation}  
gives us  the masses $m_2$ and $a\l m_2$, correspondingly.  
Thus, the naturalness condition follows from the (2,2) entry in the case of the 
inverse mass hierarchy of the flavor,  while in the quark sector this condition 
follows from the (1,1) entry.
\par    
Therefore it is necessary to clarify the concepts of naturalness for the large mixing 
angle.
Furthermore, the recent  experiments \cite{Atm1}$\sim$\cite{SKam} also indicate 
the large flavor mixing  in the neutrinos.
Let us consider the $2\times 2$ Hermitian matrix $M_\nu$.
Again, assume for simplicity that it can be diagonalized by some orthogonal 
matrix $O_\nu$:    
\begin{equation}    
O_\nu^T M_\nu O_\nu=M_\nu^{diag}\equiv     
\left(\begin{array}{cc}    
 m_1 & 0 \\    
 0   & m_2 \\    
\end{array}\right).  
\end{equation}    
Here the orthogonal matrix $O_\nu$ is     
\begin{equation}    
O_\nu=\left(\begin{array}{cc}    
c & s \\    
-s & c \\    
\end{array}\right) \ , \qquad        
c\equiv\cos\theta_\nu\simeq\frac{1}{\sqrt{2}} \ , \qquad     
s\equiv\sin\theta_\nu\simeq\frac{1}{\sqrt{2}}.    
\end{equation}    
Then $M_\nu$ is expressed in terms of the mass eigenvalues and mixing  
angle as follows:     
\begin{equation}    
M_\nu=O_\nu M_\nu^{diag} O_\nu^T=\left(\begin{array}{cc}    
m_1c^2+m_2s^2 & cs(m_2-m_1) \\    
cs(m_2-m_1) & m_1s^2+m_2c^2 \\ \end{array}\right) .    
\end{equation}    
\par    
If the neutrino masses have the hierarchy such as $m_1/m_2 =\e \ll 1$ with     
the large mixing angle,     
 the mass matrix $M_\nu$ is    
\begin{equation}    
M_\n \simeq \left(\matrix{\e c^2+s^2  & cs \cr  cs    & c^2+\e s^2 \cr}  
\right)m_2 \ .    
\end{equation}    
In the (1,1) entry, $\epsilon c^2$ should be fine tuned against $s^2$ in order to 
arrive at $m_1/m_2\sim \e$.     
For example, taking $c\simeq s\simeq 1/\sqrt{2}$, the matrix  
\begin{equation}         
 M_\nu \simeq \left (\matrix{    
               \frac{1}{2} & \frac{1}{2}\cr     
             \frac{1}{2}  &  \frac{1}{2} \cr   
                   } \right )m_2  
\label{simple}
\end{equation}  
gives $m_1 \simeq 0$ and $m_2$.
Further if we assume that $\e\simeq a\l^2$, the matrix $M_\nu$ in (\ref{simple}) 
should be replaced by the following matrix
\begin{equation}         
M_\nu\simeq \left(\matrix{    
               \frac{1+a\l^2}{2} & \frac{1}{2}\cr     
             \frac{1}{2}  &  \frac{1+a\l^2}{2} \cr   
                   } \right )m_2  \ .  
\end{equation}  
The latter gives the mass $m_1=O(\l^2)m_2$ provided that the order of $O(\l^2)$ 
is fine-tuned.  
Thus, the hierarchical neutrino masses are unnatural in the case of the  
large mixing angle.     
\par    
In the case when neutrino masses are approximately degenerate such as    
$m_1 \simeq m_2$ and $(m_2-m_1)/m_2 =\e \ll 1$, the mass matrix $M_\n$ is    
\begin{equation}    
M_\n \simeq \left (\matrix{ 1  & cs\e \cr cs\e & 1 \cr} \right)m_2 \ .    
\label{nat1}    
\end{equation}    
\noindent 
Hence, no fine tuning is required for any entry.     
We can explore the same argument when     
the mass eigenvalues are $-m_1$ and $m_2$, in which case       
\begin{equation}    
M_\nu \simeq \left(\matrix{ c^2\e  & 2 cs \cr 2 cs  &  s^2\e \cr}  
\right)m_2 \ .\label{nat2}    
\end{equation}    
Therefore  in the case of the large  mixing angle we call     
eqs.(\ref{nat1},\ref{nat2})  the natural mass matrix.    
\par  
In the sequel the naturalness of the mass matrix for the lepton sector is 
investigated for the three family model.                      
\section{Pattern of neutrino masses  and mixing}    
The present evidences in favor of neutrino masses and mixing from     
solar, atmospheric and accelerator experiments cannot be all reconciled     
in the three family framework, unless some data are excluded.        
Therefore, one should classify the possible patterns of neutrino masses     
and mixing \cite{Smi}.    
\par        
The deficit of $\nu_e$ source is measured in the solar neutrino experiment.   
The LSND experiments provide possible $\nu_e$ appearance by using $\nu_\mu$ source. 
The atmospheric neutrino experiment suggests the deficit of $\nu_\mu$ source.   
Thus in the framework of three family, the LSND results give us 
the information on $\nu_\mu \rightarrow \nu_e$oscillations, while the 
atmospheric neutrino experiment provides data on $\nu_\mu \rightarrow   
\nu_e$ or $\nu_\mu \rightarrow \nu_\tau$ oscillations. 
Also $P(\nu_e \rightarrow \nu_e$) is  measured in the solar neutrino 
experiments.      
\par     
We have the following relations for the solar neutrino \cite{solar} and the 
atmospheric neutrino \cite{Atm1}$\sim$\cite{SKam} and LSND \cite{LSND}, 
correspondingly:        
\begin{eqnarray}         
    \Delta m^2_\odot&\simeq& (0.3 \sim 1.2)\times 10^{-5} \eV^2 \nonumber\\        
                    &{\rm with}&\quad        
                         \sin^2 2\theta_{\odot}\simeq (0.65 \sim 0.85)(\rm         
                             MSW  \ large \ angle \ solution) \nonumber\\        
                    & &\quad\sin^2 2\theta_{\odot} \simeq (0.1 \sim 2)    
                         \times 10^{-2} (\rm MSW \ small \  angle \  solution) \nonumber \\  
&{\rm or}& \nonumber\\  
          &\simeq&(5 \sim 8)\times 10^{-11}\eV^2 \nonumber\\  
 &{\rm with}&\quad \sin^2 2\theta_{\odot}\simeq 0.8 \sim 1.0   
 (\rm Just-so \  solution)    
                \label{Hi}                                           \\        
    \Delta m^2_{\rm atm}&\simeq& (0.3 \sim 3)\times 10^{-2} \eV^2\qquad{\rm     
                       with}\quad         
                    \sin^2 2\theta_{atm}\simeq(0.6 \sim 1.0)   ,     
                            \nonumber\\        
    \Delta m^2_{\rm LSND}&\simeq& 0.3\sim 2 \eV^2 \qquad{\rm with}\quad         
                    \sin^2 2\theta_{LSND}\simeq(0.2 \sim 3 )\times 10^{-2}\ .    
                            \nonumber        
\end{eqnarray}         
\noindent         
In the case of three neutrinos there is an obvious relation:        
\begin{equation}         
 \Delta m^2_{21}+\Delta m^2_{32}=\Delta m^2_{31} \ ,        
\end{equation}          
\noindent        
which however, is not satisfied in the view of eq.(\ref{Hi}) unless we disregard at     
least one of the three above mentioned experimental results.        
Therefore it is impossible to reconcile all the neutrino anomalies with three 
neutrinos.        
Next  we discuss the following possibilities within the three family model for 
light neutrinos:    
\par    
\begin{description}        
\item[(a)] Choose two following solutions 
\begin{description}    
\item[(a-1)]$\Delta m^2_{\rm LSND}=\Delta     
        m^2_{\rm atm}\sim (0.2-0.3)\eV^2$\ \cite{CaFu}        
\item[(a-2)]$\Delta m^2_\odot=\Delta m^2_{\rm atm}\sim 10^{-2}\eV^2$\     
\cite{AcPa}   
\end{description}
by stretching the data.     
\item[(b)] Neglect one of the following three neutrino anomalies,
\begin{description}    
\item[(b-1)]solar neutrino data,    
\item[(b-2)]LSND experimental data,      
\item[(b-3)]atmospheric neutrino data.    
\end{description}    
\end{description}    
\par        
\subsection{Case (a)}        
The case(a-1) has been discussed by Cardall and Fuller \cite{CaFu}.        
The authors presented the masses and mixings:        
\begin{eqnarray}        
&&\Delta m^2_{21}\simeq 7\times 10^{-6} \eV^2 \ , \        
 \Delta m^2_{31}\simeq  \Delta m^2_{32} \simeq 0.3 \eV^2 \ , \nonumber \\        
&&V_1 \simeq         
 \left (\matrix{0.994 & 0.044 & 0.100 \cr -0.108 & 0.530 & 0.841 \cr        
      -0.015 & -0.847 & 0.532 \cr} \right ) \ ,        
\label{V1}
\end{eqnarray}        
\noindent     
where the MSW small angle solution of the solar neutrino is taken.     
The MSW large angle solution and "just-so" solution  can also be accommodated     
in this scheme \cite{FLM}.    
For the absolute masses one expects $m_3\gg m_2 \geq m_1$ with 
$m_3\simeq 0.5-0.6$eV.
The scenario of    $m_3\simeq m_2 \simeq m_1\simeq 1\eV$ is also possible.    
\par
Solution (\ref{V1}) requires the large values ${\cal O}(1/\sqrt{2})$ for the 
(2,3) and (3,2) entries of the mixing matrix.         
This scenario appears to be the most natural due to the oscillation evidences.
The only drawback is the behavior of  zenith angle distribution in multi-GeV 
atmospheric data in the Kamiokande experiment.        
Super-Kamiokande will be able to answer the question on the zenith angle     
dependence. 
Notice, that the preliminary analysis of Super-Kamiokande already suggested 
this dependence in \cite{SKam}.        
\par     
In the case(a-2),  the smaller neutrino mass difference     
is accounted for  both the solar and the atmospheric neutrino problems. 
The larger mass difference is under consideration for the LSND 
experiments \cite{AcPa}.         
The parameters of the mass differences and mixing are        
\begin{eqnarray}        
     &&\Delta m^2_{21}\simeq  10^{-2} \eV^2 \ , \qquad        
	 \Delta m^2_{31}\simeq  \Delta m^2_{32} \simeq (1-2) \eV^2 \ ,     
               \nonumber \\        
     &&V_2 \simeq         
	 \left (\matrix{0.700 & 0.700 & 0.140 \cr -0.714 & 0.689 & 0.124 \cr    
          -0.010 & -0.187 & 0.982 \cr} \right ) \ .        
\end{eqnarray}        
We obtain a solution $m_3\gg m_2 \geq m_1$ with $m_3\simeq {\cal O}(1\eV)$
for the absolute masses.    
This  solution  requires the large values ${\cal O}(1/\sqrt{2})$ for         
the (1,2) and (2,1) entries for the mixing matrix $V_2$.  
The atmospheric deficit of $\nu_\mu$ is interpreted by the oscillation 
$\nu_\mu \rightarrow \nu_e$.   
Note that the recent CHOOZ experiments \cite{CHOOZNew} exclude the available 
region for $\nu_\mu \rightarrow \nu_e$ oscillation given by 
Kamiokande \cite{Atm1} and Super-Kamiokande \cite{SKam}.  
Although this mixing pattern might be excluded experimentally,
we discuss its naturalness  in the next section.     
\par
\subsection{Case (b)}        
We consider the cases where one of the three neutrino     
anomalies (solar, atmospheric or LSND neutrino experiments) is neglected.
In the following the atmospheric deficit of $\nu_\mu$  is interpreted as an 
effect of $\nu_\mu \rightarrow \nu_\tau$ oscillation by taking into account 
recent CHOOZ results \cite{CHOOZNew}.    
We expel some mixing patterns  using this remarkable result.  
\vskip 0.2cm    
\noindent        
  (1){\large\it Neglecting solar neutrino (Atmospheric neutrino+LSND)}:        
\par        
In this scheme two cases are available: \\  
\noindent  
(i)$\Delta m_{31}^2\simeq\Delta m_{21}^2\simeq \Delta m^2_{\rm LSND}$ and     
$\Delta m_{32}^2 \simeq \Delta m^2_{\rm atm} \ ,$\\  
\noindent  
(ii)$\Delta m_{32}^2\simeq\Delta m_{31}^2\simeq \Delta m^2_{\rm LSND}$   
and      
$\Delta m_{21}^2\simeq \Delta m^2_{\rm atm}$.  \\
\noindent  
In the case (i), equation for the transition probability   
\begin{equation}  
P(\nu_\a \rightarrow \nu_\b)=\d_{\a\b}-4\sum_{i<j}V_{\a i}V^*_{\a j}V_{\b i}^*  
V_{\b j}\sin^2\frac{\Delta m^2_{ij}L}{4E}  
\end{equation}  
gives  
\begin{eqnarray}  
P^{\rm LSND}_{\nu_\mu \rightarrow \nu_e}&=&-4V_{\mu 1}V_{e3}V_{\mu 3}  
V_{e1}\sin^2\frac{\Delta m^2_{31}L}{4E}   
  -4V_{\mu 1}V_{e2}V_{\mu 2}V_{e1}\sin^2\frac{\Delta m^2_{21}L}{4E} \nonumber\\  
&=&4|V_{\mu 1}|^2|V_{e1}|^2\sin^2\frac{\Delta m_{\rm LSND}^2L}{4E} \ ,  
\end{eqnarray}  
and    
\begin{equation}
4|V_{\mu 1}|^2|V_{e1}|^2\simeq (0.2 \sim 3)\times10^{-2} \ ,
\end{equation}
where we use the unitarity relation   
$\sum_{i}V_{\a i}V^*_{\b i}=\d_{\a\b}$.  
The atmospheric neutrino oscillation probability  
\begin{equation}  
P^{\rm ATM}_{\nu_\mu \rightarrow \nu_\tau}=-4V_{\mu 2}V_{\tau 3}V_{\mu 3}  
V_{\tau 2}\sin^2\frac{\Delta m^2_{32}L}{4E}  
\end{equation}  
imposes $-4V_{\mu 2}V_{\tau 3}V_{\mu 3}V_{\tau 2}\simeq O(1)$.  
Thus we obtain approximately the maximal mixing  
 $|V_{\mu 2}|=|V_{\mu 3}|=|V_{\tau 2}|= |V_{\tau 3}|= 1/\sqrt{2}$.   
For the case (ii) similar analysis is applied. 
\par  
Finally this scheme needs a large mixing in $\n_\m$ and $\n_\tau$   
depending on the     
mass hierarchy pattern (i) and (ii):        
\begin{equation}   
\begin{array}{c}        m_3\simeq m_2\simeq 1 \eV \gg m_1:\\    
\end{array} \qquad V_3 \simeq \left (\matrix{1 & \e_1 & \e_2 \cr \e_3 & c & s \cr\e_4 & -s & c \cr} \right ) \ ,        
\label{V3}    
\end{equation}        
$$(\Delta m_{31}^2\simeq\Delta m_{21}^2\simeq \Delta m^2_{\rm LSND}, \     
\Delta m_{32}^2        
\simeq \Delta m^2_{\rm atm})$$         
\noindent     
or     
\begin{equation}        m_3\simeq 1\eV\gg m_2\simeq 0.1 \eV \geq m_1: \qquad   
      V_4\simeq \left (\matrix{   \e_1 & \e_2 & 1 \cr    c & s & \e_3 \cr      
	     -s & c & \e_4 \cr   } \right ) \ ,        
\label{V4}    
\end{equation}         
$$(\Delta m_{32}^2\simeq\Delta m_{31}^2\simeq \Delta m^2_{\rm LSND}, \     
\Delta m_{21}^2        
\simeq \Delta m^2_{\rm atm})$$         
\noindent 
respectively,        
where $c\equiv\cos\th\simeq 1/\sqrt{2}$ and $s\equiv \sin\th\simeq     
1/\sqrt{2}$,        
and $\e_3\simeq \l^2\sim \l^3$ due to the LSND data 
$\sin^2 \th_{LSND}\sim 10^{-3}$.        
Others $\e_i$'s are small, but not constrained at present.        
\vskip 0.2cm    
\noindent   
(2){\large\it Neglecting LSND(Solar neutrino+Atmospheric neutrino)}:        
\par    
In order to interpret the solar neutrino data, the component $V_{e3}$ should 
be taken much smaller than 1.
However, this is not compatible with the only possible solution 
$\nu_\mu\rightarrow \nu_e$ given  by the atmospheric neutrino experiments. 
This solution is also excluded by the CHOOZ 
experiment \cite{CHOOZNew} as mentioned above.  
Therefore, in the atmospheric neutrino deficit we  can take either    
$m_3\simeq 0.1\eV$  or the almost degenerate three neutrino masses     
$m_3 \simeq m_2 \simeq m_1\simeq 1{\rm eV}$.     
For the MSW small angle solution the mixing matrix will be         
\begin{equation}   
  \begin{array}{c} m_3\simeq 0.1\eV\gg m_2\simeq 10^{-3}\eV 
          \geq m_1: \\   {\rm or}\  m_1\simeq m_2\simeq m_3\simeq 1\eV:\\   
  \end{array}  \quad      	
   V_5\simeq \left (\matrix{1 & \e_1 & \e_2 \cr 
                            \e_3 & c & s \cr   
                            \e_4 & -s & c \cr} \right ) \ ,       
\label{V5}
\end{equation}       
$$(\Delta m_{32}^2\simeq\Delta m_{31}^2\simeq \Delta m^2_{\rm atm}, \  
   \Delta m_{21}^2\simeq \Delta m^2_\odot) \ ,$$         
\noindent     
with $\e_1\simeq \l^2$, which is due to 
$\sin^2 2\theta_{\odot} \simeq  10^{-2}$.     
Another solution is     
\begin{equation}         
 V_6\simeq \left (\matrix{ \e_1 & 1 & \e_2 \cr  
                           c & \e_3 & s \cr  
                           -s & \e_4 & c \cr} \right ) \ ,        
\label{V6}
\end{equation}         
\noindent        
where either of the following three cases: 
\begin{eqnarray}
m_3&\simeq&  m_2\gg m_1(\Delta m_{31}^2\simeq\Delta m_{21}^2\simeq 
\Delta m^2_{\rm atm}, \Delta m_{32}^2 \simeq \Delta m^2_\odot) \ ,\nonumber\\
m_3&\gg& m_2 \geq m_1(\Delta m_{32}^2\simeq\Delta m_{31}^2\simeq \Delta m^2_{\rm atm},  \Delta m_{21}^2  \simeq \Delta m^2_\odot)\ ,\\ 
m_1&\simeq& m_2\simeq m_3\simeq 1\eV \nonumber
\end{eqnarray} 
is allowed.   
\par
On the other hand, a typical mixing matrix pattern in the case of the just-so 
solution and the MSW large angle solution has the following form:
\begin{equation}  
\left (\matrix{ c & s & \e_1 \cr  -s & c & \e_2 \cr \e_3 & \e_4 & 1 \cr} \right )   
\quad {\rm or}\quad  
\left (\matrix{ c & s & \e_1 \cr \e_2 & \e_3 & 1 \cr -s & c & \e_4 \cr} \right ).  
\end{equation}   
However, neither of these matrices is compatible with the maximal mixing pattern 
of the atmospheric $\nu_\mu \rightarrow \nu_\tau$ oscillation. 
Of course  one can find a mixing matrix which is consistent with both large 
angles of the solar neutrino and the atmospheric neutrino within the 
experimental error bars \cite{Giunti}.  
But again, this case is too complicated for the analysis of the naturalness 
of the mixing matrix.  
We shall not discuss it in the following.
\par  
Thus the MSW small angle solution in the solar neutrino  and the atmospheric 
$\nu_\mu \rightarrow \nu_\tau$ oscillation solution are compatible with the mixing matrix $V_{5,6}$.  
We investigate the naturalness of these two mixing patterns in section 4.
\vskip 0.2cm       
\noindent   (3){\large\it Neglecting atmospheric neutrino(Solar neutrino+LSND)}:        
\par    
The standard scenario is characterized by either the strong mass hierarchy     
$m_3\gg m_2 \gg m_1$,  or the $m_2-m_1$ degenerate case   
where $m_3 \gg m_2 \simeq m_1    
(\Delta m_{32}^2\simeq\Delta m_{31}^2\simeq   \Delta m^2_{\rm LSND}, \Delta m_{21}^2        
\simeq \Delta m^2_\odot)$.        
Then the  possible mixing matrices are        
\begin{eqnarray}    
      &&V_7\simeq \left (\matrix{1 & \e_1 & \e_2 \cr \e_3 & 1 & \e_4 \cr    
	       \e_5 & \e_6 & 1 \cr} \right ) \ , \qquad       
		    	V_8\simeq \left (\matrix{\e_1 & 1 & \e_2 \cr 1 & \e_3 & \e_4 \cr   
		      \e_5 & \e_6 & 1 \cr} \right ) \ , \nonumber \\      
			      &&V_9\simeq \left (\matrix{\e_1 & 1 & \e_2 \cr \e_3 & \e_4 & 1 \cr    
				       1 & \e_5 & \e_6 \cr} \right ) \ , \qquad        
					   	V_{10}\simeq \left (\matrix{1 & \e_1 & \e_2 \cr \e_3 & \e_4 & 1 \cr    
						     \e_5 & 1 & \e_6 \cr} \right ) \ .         
\label{V78910}
\end{eqnarray}         
\noindent  
for the MSW small mixing solution in the solar neutrino experiments.  
Here  we have $\e_2\simeq \l$, $\e_4\simeq  \l^{3/2}$ for $V_7$ and $V_8$  
 where $\e_2\e_4$ and $\e_4$  are  constrained from the LSND data($\n_\m-\n_e$)  
  and CHORUS/NOMAD data($\n_\m-\n_\tau$), respectively.  
The relation $\e_2\simeq \l^2\sim \l^3$ for $V_9$ and $V_{10}$ is given by  the LSND data.    
The  mixing matrices $V$     
with the (1,3) entry,  being nearly 1,      
are not allowed by the solar neutrino deficit \cite{Mina}$\sim$\cite{tani}.    \par  
On the other hand, taking the just-so or the MSW large angle solutions, 
we obtain the following matrices   
\begin{equation}  
V_{11}=\left (\matrix{ c &s & \e_1 \cr   -s & c & \e_2 \cr  \e_3 & \e_4 & 1 \cr} \right )   
\quad {\rm or}\quad  
V_{12}=\left (\matrix{ c &s & \e_1 \cr  \e_2 & \e_3 & 1 \cr   -s &c & \e_4 \cr } \right ),  
\label{V11}
\end{equation}  
where  $\e_1 \simeq \l$ and $\e_2 \simeq \l^{3/2}$ for $V_{11}$ and  
 $\e_1 \simeq \l^2 \sim \l^3$ for $V_{12}$ due to  the LSND data and CHORUS/NOMAD data.  
Notice,  that matrices in (\ref{V11}) require small values for $V_{e3}V_{\mu 3}$.  
\section{Natural lepton mass matrix}         
We discuss here the mass matrices of the charged lepton $M_E$ and  the light      
Majorana neutrino   $M_{LL}$, in which $CP$ violation is neglected.        
Let us begin with  the charged lepton sector.        
The mass matrix $M_E$ is diagonalized by the orthogonal matrix $O_E$, which        
is parameterized in terms of three angles:        
\begin{equation}          
  O_E = \left (\matrix{ c_{13} c_{12} & c_{13} s_{12} &  s_{13} \cr         
  -c_{23}s_{12}-s_{23}s_{13}c_{12} & c_{23}c_{12}-s_{23}s_{13}s_{12} &         
                       s_{23}c_{13} \cr        
  s_{23}s_{12}-c_{23}s_{13}c_{12} & -s_{23}c_{12}-c_{23}s_{13}s_{12} &         
                       c_{23}c_{13} \cr} \right ) \ ,        
\end{equation}         
\noindent       
where  $s_{ij}\equiv \sin{\th_{ij}}$ and $c_{ij}\equiv \cos{\th_{ij}}$ are      
mixings, and  the $CP$ violating phase is neglected.          
\par        
For the charged leptons, the mass hierarchy is clearly determined as follows:        
\begin{equation}         
      {m_e\o m_\m} \simeq 2\times \l^4 \ , \qquad {m_\m\o m_\tau}     
           \simeq \l^2 \ .        
\end{equation}        
\noindent        
Choosing properly an orthogonal matrix,        
we  can satisfy the relation $\th_{ij} \ll 1$ due to  the large     
mass hierarchy.  							    
Next applying the same naturalness arguments as in section 2, we get      
$\th_{12}\leq \l^2$,  just as in the case of the up-quark sector.        
We derive $\th_{23}\leq \l$ and $\th_{13}\leq \l^3$ for the rest possible angles.        
Finally, we obtain the orthogonal matrix which diagonalizes $M_E$:        
\begin{equation}          
  |O_{E}| \simeq  \left (\matrix{1& \leq\l^2 & \leq\l^3 \cr        
        \leq\l^2 &1  & \leq\l\cr    \leq \l^3 & \leq\l& 1 \cr} \right ) \ .    
\end{equation}        
\par          
Using this orthogonal matrix $O_E$, we can study now the natural neutrino mass     
matrix.        
Recall, that there are some expected patterns of neutrino masses and mixing(see section 3). 
We shall investigate the naturalness of those patterns.        
Our strategy will be the following.        
At the first stage  we construct the light Majorana neutrino mass matrix
\begin{equation}          
    M_{LL}=O^T_{\n} M^{\rm diag}_{LL} O_{\n} \ .        
\label{MLL}    
\end{equation}        
\noindent        
Matrix $O_\n$ is obtained from the relation $V=O^T_{E} O_{\n}$.
Note, that various  $V$'s were discussed  in section 3. 
More precisely, 
there are the cases (a) and (b)         
under the assumption of the $CP$ conservation in the lepton sector.        
Thus, we can argue now the naturalness of $M_{LL}$.        
If the relevant  $M_{LL}$ is a natural mass matrix at this stage,        
we refer to  this  as the first stage natural mass matrix.         
Then  we proceed to the second stage described below  in section 5.        
\par
In the see-saw mechanism \cite{see}, one obtains $M_{LL}$ as
\begin{equation}          
    M_{LL}=M_{LR} M_{RR}^{-1} M_{LR}^T \ ,        
\end{equation}        
where $M_{LR}$ is the Dirac mass matrix and $M_{RR}$ is the right-handed     
Majorana mass matrix.          
We argue the naturalness of the  $M_{RR}$  leading to $M_{LL}$ using the 
hierarchical natural mass matrix $M_{LR}$, obtained        
 by Peccei and Wang \cite{Natu} for the up and the down quark sectors.
If there is no fine-tuning in $M_{RR}$,        
we call $M_{LL}$ the second stage natural matrix.       
\par        
\subsection{Case(a)}    
Consider the Cardall and Fuller mixing matrix $V_1$.        
The orthogonal matrix $O_\n$ is expected to be        
\begin{equation}          
     O_\n \simeq \left (\matrix{1 & \e_1 & \e_2 \cr \e_3 & c & s\cr        
        \e_4 & -s & c \cr} \right ) \ ,        
\end{equation}        
where $\e_1\sim \l^2$, $\e_2\sim \l$, $\e_3\sim -\l$, $\e_4\sim \l^2$        
and $c\equiv\cos\th\simeq 0.5$ according to $O_\n=O_E V$ with a fixed  
natural orthogonal matrix    
\begin{equation}    
  O_{E}=  \left (\matrix{    
             1& \l^2 & -\l^3 \cr        
             -\l^2 &1  & \l^2\cr        
             \l^3 & -\l^2& 1 \cr    
                        } \right ) \ .        
\end{equation}        
Then constructed $M_{LL}$ can be rewritten as follows:        
\begin{equation}          
M_{LL}^{(1)} \simeq \left (\matrix{m_1+m_2\e_3^2+m_3\e_4^2 & m_1\e_1+m_2c\e_3 -m_3 s\e_4         
  & m_1\e_2+m_2 s\e_3+m_3 c\e_4 \cr         
  m_1\e_1+m_2c\e_3 -m_3 s\e_4  & m_1\e_1^2+m_2 c^2 +m_3 s^2         
  & m_1\e_1\e_2+cs(m_2-m_3)\cr        
  m_1\e_2+m_2 s\e_3+m_3 c\e_4 &  m_1\e_1\e_2+cs(m_2-m_3)         
& m_1\e_2^2+m_2 s^2+m_3 c^2 \cr} \right ) \ .        
\label{MLL1}        
\end{equation}        
    
\noindent   
The naturalness condition of the $3\times 3$ mass matrix in eq.(\ref{MLL1}) 
is obtained from the following inspection.  
The $m_1$, $m_2$ and $m_3$ terms  with  coefficients ${\cal O}(1)$ on the diagonal entries 
are essential ones for the mass eigenvalues $m_1$, $m_2$ and $m_3$ as we have seen 
in section 2.  
Hence, we cannot neglect these terms. 
 Then  the naturalness  of the matrix is easily found  focusing
   on $m_1$, $m_2$ and $m_3$ terms  with  coefficients 
${\cal O}(1)$ on the diagonal entries.   
In order to avoid the fine-tuning of (1,1), (2,2) and (3,3) entries  in $M_{LL}$,        
we find the conditions        
\begin{equation}          
m_1\geq m_2\e_3^2+m_3\e_4^2 \ , \quad m_2 c^2 \geq m_1\e_1^2+m_3 s^2  \ ,    
\quad m_3c^2\geq m_1\e_2^2+m_2s^2 \ ,      
 \label{CMLL1}      
\end{equation}        
\noindent         
which are satisfied as long  as  $m_3 \simeq m_2$ due to $c\simeq 0.5$.    
Taking into consideration the relations  
$\Delta m^2_{21}\simeq 7 \times 10^{-6} \eV^2$    
and  $\Delta m^2_{31}\simeq  \Delta m^2_{32} \simeq 0.3 \eV^2$         
 in the Cardall and Fuller scheme, we obtain     
\begin{equation}          
m_3 \simeq m_2 \simeq m_1 \ .        
\end{equation}        
         
   In other words, the hierarchical neutrino masses are excluded by         
the principle of the  naturalness of the mass matrix.        
The three neutrino masses should be almost degenerated.        
This result is still valid for a few variations of the mixing matrix \cite{FLM}.     
\par      
In the case (a-2) where 
$\Delta m^2_\odot=\Delta m^2_{\rm atm}$ is required \cite{AcPa},     
the analysis is similar to the previous case. The orthogonal matrix         
$O_\nu=O_EV_2$ is        
\begin{equation}          
O_\n \simeq \left (\matrix{c & s & \e_1 \cr -s & c & \e_2 \cr        
        \e_3 & \e_4 & 1 \cr} \right ) \ ,        
\end{equation}        
where $\e_1\sim \e_2 \sim -\e_3 \sim -\e_4 \sim \l$        
and $c\simeq s \simeq 1/\sqrt{2}$.        
The $M_{LL}$ is     
\begin{equation}          
M_{LL}^{(2)} \simeq \left (\matrix{m_1c^2+m_2s^2+m_3\e_3^2         
                           & (m_1-m_2)cs+m_3c\e_3\e_4        
                           & m_1c\e_1-m_2 s\e_2+m_3 \e_3 \cr        
                             (m_1-m_2)cs+m_3c\e_3\e_4         
                           & m_1 s^2+m_2 c^2 +m_3 \e_4^2        
                           & m_1 s\e_1+m_2 c\e_2+m_3\e_4 \cr        
                              m_1c\e_1-m_2 s\e_2+m_3 \e_3         
                           & m_1 s\e_1+m_2 c\e_2+m_3\e_4        
                           & m_1\e_1^2+m_2\e_2^2+m_3 \cr} \right ) \ ,        
\label{MLL2}    
\end{equation}     
derived from eq.(\ref{MLL}).    
Using this matrix, we can put the naturalness condition for (1,1), (2,2) and (3,3)    
 entries as        
\begin{equation}      
        m_1c^2\geq m_2s^2+m_3\e_3^2\ ,     
      \quad m_2 c^2 \geq m_1s^2+m_3\e_3^2 \ ,     
      \quad m_3\geq m_1\e_1^2+m_2\e_2^2 \ ,        
\end{equation}        
which leads to    
\begin{equation}          
 m_3 \gg m_2 \simeq m_1 \qquad  {\rm or}\qquad  m_3 \simeq m_2 \simeq m_1 \ .        
\end{equation}        
\noindent    
The first case of the mass patterns require 
the condition $m_1/m_3\simeq m_2/m_3 \sim \e_3^2\sim \l^2$. 
Thus  $V_2$ could be a natural one.        
The latter case requires the neutrino masses to be much larger than $1\eV$    
in order to reproduce $\Delta m^2_{21}\simeq  10^{-2} \eV^2$ and    
$\Delta m^2_{31}\simeq \Delta m^2_{32} \simeq (1-2) \eV^2$.    
However, these masses contradict to the neutrinoless double beta decay 
experiments \cite{DBe}.    
Therefore  we do not consider this case  in our paper.    
\par    
Table 1 summarizes the results on the     
naturalness for $V_1$ and $V_2$.    
\begin{center}
\framebox{Table 1}
\end{center}
\par    
\subsection{Case(b)}    
\vskip 0.2cm    
\noindent        
(1){\large\it Neglecting solar neutrino (Atmospheric neutrino+LSND)}:        
\par
In the case of $V_3$ in eq.(\ref{V3}), the orthogonal matrix $O_\n$ is expected to be
\begin{equation}          
     O_\n \simeq \left (\matrix{1 & \e_1 & \e_2 \cr \e_3 & c & s\cr        
        \e_4 & -s & c \cr} \right ) \ ,        
\end{equation}        
where $\e_3\sim \l^2$  and $c\simeq s\simeq 1/\sqrt{2}$.        
The neutrino masses satisfy the hierarchy $m_3\simeq m_2 \gg m_1$.        
The matrix $M_{LL}^{(3)}$ is the same as $M_{LL}^{(1)}$     
in eq.(\ref{MLL1}),         
and the conditions of naturalness are the same as in eq.(\ref{CMLL1}).        
This is consistent with the hierarchy 
$m_3\simeq m_2\gg m_1$ assuming  
the condition $m_1\geq m_2\e_3^2+m_3\e_4^2$.        
Therefore, $V_3$ leads to  a natural mass matrix.    
\par      
In the case of $V_4$ in eq.(\ref{V4}), the orthogonal matrix $O_\n$ is expected to be 
\begin{equation}          
     O_\n \simeq \left (\matrix{    
                  \e_1 & \e_2 & 1 \cr     
                     c & s & \e_3 \cr        
                    -s & c & \e_4 \cr    
                               } \right ) \ ,        
\end{equation}        
\noindent     
with $m_3\gg m_2 \geq m_1$.        
Then,  $M_{LL}$ is given by
\begin{equation}          
     M_{LL}^{(4)} \simeq \left (\matrix{    
                    m_1\e_1^2+m_2c^2+m_3s^2           &     
                    m_1\e_1\e_2+(m_2 -m_3)cs          &     
                    m_1\e_1+m_2c\e_3 -m_3 s\e_4       \cr         
	            m_1\e_1\e_2+(m_2 -m_3)cs          &     
                    m_1\e_2^2+m_2 s^2 +m_3 c^2        &     
                    m_1\e_2+m_2 s \e_3 +m_3c\e_4      \cr        
                    m_1\e_1+m_2c\e_3-m_3s\e_4         &      
                    m_1\e_2+m_2 s\e_3+m_3 c \e_4      &     
                    m_1+m_2 \e_3^2+m_3 \e_4^2         \cr    
                                        } \right ) \ .        
\end{equation}        
We derive the conditions of naturalness:        
\begin{equation}          
    m_2c^2\geq m_1\e_1^2  +m_3s^2 \ , \quad     
    m_3 c^2 \geq m_1\e_2^2+m_2 s^2 \ , \quad        
    m_1\geq  m_2 \e_3^2+m_3 \e_4^2\ .        
\end{equation}        
\noindent     
These require $m_3\simeq m_2\gg m_1$,  which contradicts to    
$m_3\gg m_2\geq m_1$ determined by eq.(\ref{V4}).    
Thus, $V_4$ leads to an unnatural mass matrix.        
\par
\vskip 0.2cm    
\noindent
(2){\large\it Neglecting LSND(Solar neutrino+Atmospheric neutrino)}:        
\par
In the case of $V_5$ in eq.(\ref{V5}), the orthogonal matrix $O_\n$ is expected to be        
\begin{equation}          
     O_\n \simeq \left (\matrix{1 & \e_1 & \e_2 \cr \e_3 & c & s\cr        
        \e_4 & -s & c \cr} \right ) \ ,        
\end{equation}        
where $\e_1\sim \l^2$  and $c\simeq s\simeq 1/\sqrt{2}$.        
The neutrino masses satisfy either the hierarchy $m_3\gg m_2 \geq m_1$  or    
    $m_3 \simeq m_2 \simeq m_1$.        
The $M_{LL}^{(5)}$ is the same as $M_{LL}^{(1)}$     
in eq.(\ref{MLL1}).        
Then  the conditions of the natural mass matrix are also the same as in 
eq.(\ref{CMLL1}).    
This is in consistence  with the case of    
$m_3 \simeq m_2 \simeq m_1$.       
Therefore, $V_5$ leads to a natural mixing matrix for the degenerate masses,    
 but does not for  $m_3\gg m_2 \geq m_1$.       
\par      
In the case of $V_6$ in eq.(\ref{V6}), the orthogonal matrix $O_\n$ is 
\begin{equation}          
     O_\n \simeq \left (\matrix{    
                  \e_1 & 1 & \e_2 \cr     
                   c & \e_3 & s \cr        
                   -s & \e_4 & c \cr    
                         } \right ) \ ,        
\end{equation}        
where $\e_1\sim \l^2$ and $c\simeq s\simeq 1/\sqrt{2}$        
with $m_3\simeq  m_2 \gg m_1$.        
Then  $M_{LL}$ is given by        
\begin{equation}          
     M_{LL}^{(6)} \simeq \left (\matrix{m_1\e_1^2+m_2c^2+m_3s^2     
          & m_1\e_1+m_2c\e_3 -m_3 s\e_4         
	  & m_1\e_1\e_2+(m_2 -m_3)cs \cr         
	  m_1\e_1+m_2c\e_3 -m_3 s\e_4  & m_1+m_2 \e_3^2 +m_3 \e_4^2         
	   & m_1\e_2+m_2 s \e_3 +m_3c\e_4\cr        
    m_1\e_1\e_2+(m_2-m_3)cs &  m_1\e_2+m_2 s\e_3+m_3 c \e_4         
	& m_1\e_2^2+m_2 s^2+m_3 c^2 \cr} \right ) \ .        
\label{MLL6}        
\end{equation}        
Considering the (2,2) entry of the matrix, we conclude that the natural 
mass matrix is defined correctly if        
\begin{equation}          
    m_1\geq m_2 \e_3^2+m_3 \e_4^2 \ .    
\end{equation}        
\noindent     
The latter is satisfied for the both cases: $m_3\simeq  m_2 \gg m_1$ and $m_3\simeq  m_2 \simeq m_1$.        
Thus $V_6$ leads to a natural mass matrix for both $m_3\simeq  m_2 \gg m_1$    
 and $m_3\simeq  m_2 \simeq m_1$.        
\par      
\vskip 0.2cm       
\noindent
(3){\large\it Neglecting the atmospheric neutrino(Solar neutrino+LSND)}:      
\par    
The standard scenario is characterized by the strong mass hierarchy     
$m_3\gg m_2 \geq m_1$.        
For the case of $V_7$ in eq.(\ref{V78910}),   we obtain the constrains  from the solar neutrino     
and the LSND data:        
 $\e_1 \leq \l^2$, $\e_2\sim \l$, $\e_4\leq \l^{3/2}$. Then,   the $M_{LL}$ matrix is         
\begin{equation}          
      M_{LL}^{(7)} \simeq \left (\matrix{m_1+m_2\e_3^2+m_3\e_5^2     
            & m_1\e_1+m_2\e_3 +m_3 \e_5\e_6         
	  & m_1\e_2+m_2 \e_3\e_4+m_3 \e_5 \cr         
	  m_1\e_1+m_2 \e_3 +m_3 \e_5\e_6  & m_1\e_1^2+m_2 +m_3 \e_6^2         
	   & m_1\e_1\e_2+m_2\e_4+m_3\e_6\cr        
     m_1\e_2+m_2 \e_3\e_4+m_3 \e_5 &  m_1\e_1\e_2+m_2\e_4+m_3\e_6        
	 & m_1\e_2^2+m_2 \e_4^2+m_3 \cr} \right ) \ .         
\end{equation}        
\noindent     
In order to avoid the fine-tuning, one should impose  the following 
conditions for the (1,1) and (2,2) entries:        
\begin{equation}          
    m_1\geq m_2 \e_3^2 + m_3 \e_5^2  \ , \qquad     
    m_2 \geq m_1\e_1^2+m_3\e_6^2 \ .        
\end{equation}        
\noindent        
Choosing  $\e_3\leq \l^2$, $\e_5\sim \l$ and $\e_6\leq\l^2$ we satisfy  
these conditions in the case where  
$m_1\simeq m_2\l \simeq m_3\l^2$. 
However this hierarchy is not sufficient to ensure  
$\Delta m_{32}^2\simeq \Delta m_{31}^2 \simeq \Delta m^2_{LSND}$
and $\Delta m_{21}^2\simeq \Delta m^2_\odot$.
But the latter  contradicts to the mass hierarchy $m_3\gg m_2\gg m_1$.
Thus  $V_7$ leads to  an unnatural mass matrix.        
\par    
For $V_8$ in eq.(\ref{V78910}) we obtain          
\begin{equation}          
     M_{LL}^{(8)} \simeq \left (\matrix{m_1\e_1^2+m_2 +m_3\e_5^2 &     
          m_1\e_1+m_2 \e_3 +m_3 \e_5\e_6         
	  & m_1\e_1\e_2+m_2\e_4 +m_3\e_5 \cr         
	  m_1\e_1+m_2\e_3 +m_3 \e_5\e_6  & m_1+m_2 \e_3^2 +m_3 \e_6^2         
	   & m_1\e_2+m_2 \e_3\e_4 +m_3\e_6\cr        
    m_1\e_1\e_2+m_2\e_4+m_3\e_5 &  m_1\e_2+m_2 \e_3\e_4+m_3\e_6         
	& m_1\e_2^2+m_2 \e_4^2+m_3 \cr} \right ) \ ,        
\end{equation}        
\noindent        
which leads to  an unnatural mass matrix.
Indeed,   the naturalness conditions        
\begin{equation}          
    m_2\geq m_1\e_1^2 + m_3 \e_5^2  \ , \quad     
    m_1\geq m_2\e_3^2 +m_3\e_6^2 \ ,        
\end{equation}        
\noindent     
cannot be satisfied for the mass hierarchy $m_3\gg m_2\geq m_1$ according
to the similar argument in the previous case.        
Thus, we find that  $V_8$ is unnatural as well as $V_7$.     
\par           
For $V_9$ in eq.(\ref{V78910}), we derive         
\begin{equation}          
     M_{LL}^{(9)} \simeq \left (\matrix{m_1\e_1^2+m_2\e_3^2 +m_3     
          & m_1\e_1+m_2 \e_3\e_4 +m_3 \e_5        
	  & m_1\e_1\e_2+m_2\e_3+m_3\e_6 \cr         
	  m_1\e_1+m_2\e_3\e_4 +m_3 \e_5  & m_1+m_2 \e_4^2 +m_3 \e_5^2         
	   & m_1\e_2+m_2\e_4 +m_3\e_5\e_6\cr        
    m_1\e_1\e_2+m_2\e_3+m_3\e_6 &  m_1\e_2+m_2\e_4+m_3\e_5\e_6         
	& m_1\e_2^2+m_2+m_3\e_6^2 \cr} \right ) \ ,        
\label{MLL9}           
\end{equation}        
\noindent        
 and the corresponding naturalness conditions        
\begin{equation}          
    m_2\geq m_1\e_2^2+m_3 \e_6^2  \ , \quad m_1\geq m_2\e_4^2+m_3\e_5^2 \ .        
\label{M9}       
\end{equation}
These conditions are similar to the cases of $V_{7,8}$,  and we can choose 
arbitrary small values $\e_5$ and $\e_6$, keeping $\e_6\geq \e_5$.
For example, as in the case $m_3 \gg m_2 \simeq m_1$,  we obtain 
the mixing matrix 
\begin{equation}
V_{9}\simeq\left(\matrix{
      -\l^2 & 1 & \l^3 \cr
       -\l^2 & -\l^3 & 1 \cr
         1  & \l^2 & \l^2 \cr}\right) \ ,
\label{M9dg}
\end{equation}
which is consistent with the condition(\ref{M9}). 
Here we input $\e_2\simeq \l^3, \e_5 \simeq \e_6 \simeq \l^2$,   
and the masses 
$m_3\gg m_2 \simeq m_1\simeq m_3\l^4$.
Another choice is the case $\e_6 \gg \e_5$ which corresponds to 
$m_3 \gg m_2 \gg m_1$. 
As a typical case, we take $\e_6\simeq \l^2$ and $\e_5\simeq \l^4$.
Then the mixing matrix becomes
\begin{equation}
V_{9}\simeq\left(\matrix{
      -\l^4 & 1 & \l^3 \cr
       -\l^2 & -\l^3 & 1 \cr
         1  & \l^4 & \l^2 \cr}\right) \ ,
\label{M9hie}
\end{equation}
with $m_3\gg m_2 \gg m_1(m_2 \simeq m_3\l^4, m_1\simeq m_3\l^8)$.
Both  cases are consistent with 
$\Delta m_{32}^2\simeq \Delta m_{31}^2 \simeq \Delta m^2_{LSND}$
and $\Delta m_{21}^2\simeq \Delta m^2_\odot$.        
Thus, $V_9$ could be a natural mass matrix for $m_3 \gg m_2 \gg m_1$ and 
$m_3 \gg m_2 \simeq m_1$.        
\par      
For $V_{10}$ in eq.(\ref{V78910})  we obtain 
\begin{equation}          
      M_{LL}^{(10)} \simeq \left (\matrix{m_1+m_2\e_3^2+m_3\e_5^2 &     
          m_1\e_1+m_2\e_3\e_4 +m_3 \e_5         
	  & m_1\e_2+m_2 \e_3+m_3 \e_5\e_6 \cr         
	  m_1\e_1+m_2 \e_3\e_4 +m_3 \e_5  & m_1\e_1^2+m_2\e_4^2 +m_3         
	   & m_1\e_1\e_2+m_2\e_4+m_3\e_6\cr        
     m_1\e_2+m_2 \e_3+m_3 \e_5\e_6 &  m_1\e_1\e_2+m_2\e_4+m_3\e_6        
	 & m_1\e_2^2+m_2+m_3\e_6^2 \cr} \right ) \ ,         
\end{equation}        
\noindent 
with the naturalness conditions      
\begin{equation}          
    m_2\geq m_1\e_2^2+m_3 \e_6^2  \ , \quad m_1\geq m_2\e_3^2+m_3\e_5^2 \ .        
\end{equation}        
These relations are held as long as $\e_6\geq \e_5$.
According to the similar analysis given in the previous case($V_9$),         
$V_{10}$ can yield a natural mass matrix.        
\par      
For $V_{11}$ in eq.(\ref{V11}), we obtain
\begin{equation}          
      M_{LL}^{(11)} \simeq \left (\matrix{m_1c^2+m_2 s^2+m_3\e_3^2 &     
          (m_1-m_2)cs+m_3\e_3\e_4         
	  & m_1\e_1c-m_2 \e_2s+m_3 \e_3 \cr         
	  (m_1-m_2)cs+m_3\e_3\e_4  & m_1s^2+m_2c^2 +m_3\e_4^2         
	   & m_1\e_1s+m_2\e_2c+m_3\e_4\cr        
     m_1\e_1c-m_2 \e_2s+m_3 \e_3 &  m_1\e_1s+m_2\e_2c+m_3\e_4       
	 & m_1\e_1^2+m_2\e_2^2+m_3 \cr} \right ) \ ,         
\label{V11mass}  
\end{equation}        
\noindent 
with the naturalness conditions
\begin{equation}          
    m_1c^2 \geq m_2s^2 + m_3\e_3^2  \ , \quad m_2 c^2 \geq m_1 s^2 + m_3\e_3^2 \ ,        
\end{equation}        
where  $c\simeq s \simeq \frac{1}{\sqrt{2}}$.  
This implies 
\begin{equation}  
  m_1\simeq m_2 \ , \qquad \frac{m_2}{m_3}\geq \e_3^2 \ ,  
\end{equation}  
{\it i.e.} the conditions, under which  the mixing matrix $V_{11}$ could 
lead to a natural mass matrix, but only in the case when 
$m_3 \gg m_2\simeq m_1$.     
\par  
For $V_{12}$ in eq.(\ref{V11}), we obtain  
\begin{equation}          
      M_{LL}^{(12)} \simeq \left (\matrix{m_1c^2+m_2\e_2^2+m_3s^2 &     
          (m_1-m_3)cs+m_2\e_2\e_3         
	  & m_1\e_1c+m_2\e_2-m_3\e_4s \cr         
	  (m_1-m_3)cs+m_2\e_2\e_3   & m_1s^2+m_2\e_3^2 +m_3c^2         
	   & m_1\e_1s+m_2\e_3+m_3\e_4c \cr        
     m_1\e_1c+m_2\e_2-m_3\e_4s &  m_1\e_1s+m_2\e_3+m_3\e_4c       
	 & m_1\e_1^2+m_2+m_3\e_4^2 \cr} \right ) \ ,         
\label{V12mass}  
\end{equation}        
\noindent 
where the naturalness condition  is    
\begin{equation}          
    m_1c^2 \geq m_2 \e_2^2 + m_3 s^2  \ , \quad m_2  \geq m_1 \e_1^2 + m_3\e_4^2 \ ,        
\end{equation}  
\noindent which gives      
\begin{equation}          
      m_1\simeq m_2 \ , \qquad \frac{m_2}{m_3}\geq \e_4^2 \ .        
\end{equation}        
The mixing matrix $V_{12}$ could also give a natural mass matrix  for the   
case of $m_3\gg m_2\simeq m_1$.   
\par    
Thus in the case (b)  we obtain all together six natural mixing patterns
($V_{3,6,9,10,11,12}$) for the  hierarchical neutrino masses,    
 and two mixing patterns($V_{5,6}$) for the degenerate masses.    
The possible solutions for the neutrino masses     
from the naturalness point of view are summarized in Table2.    
\begin{center}
\framebox{Table 2}
\end{center}
\par    
In conclusion, we underline that the natural mass matrices with the hierarchical mixing patterns     
are $V_2$, $V_3$, $V_6$, $V_9$, $V_{10}$, $V_{11}$ and $V_{12}$.     
The rest of the patterns, {\it i.e.} $V_1$, $V_4$, $V_5$, $V_7$ and $V_8$, 
 lead to the unnatural mass matrices     
in case of the hierarchical mass matrices.    
However,  if the degenerate mass matrices are allowed as the natural     
mass matrices, the mixing solutions       
$V_1$, $V_5$ and $V_6$ lead to the natural mass matrices    
at the level of first stage of naturalness.  	        
In the next section, we consider the naturalness of the        
 $M_{RR}$ for the selected $V_1$, $V_2$, $V_3$, $V_5$, $V_6$, $V_9$,   
$V_{10}$, $V_{11}$  and  $V_{12}$   
patterns.    
Here $V_1$  is a typical case with the degenerate    
neutrino masses under condition of naturalness.     
\section{Natural see-saw realization} 	        
In this section, we study the natural realization of the neutrino   
mass matrix        
 in the see-saw mechanism \cite{see}.      
Without loss of generality we consider $V_1$  as a 
sample of the three degenerate neutrino masses.
\par    
Since the mixing patterns $V_1$, $V_2$, $V_3$, $V_5$  and $V_6$        
 explain the atmospheric neutrino deficit, the large mixing angle    
 (maximal mixing)  is inevitable.     
Moreover, at least two neutrinos are     
almost degenerate in these cases.        
Our aim is to derive the large mixing angle and the degenerate masses from 
the see-saw mechanism using the hierarchical  Dirac mass matrix $M_{LR}$.      In general, there are some  conditions to reproduce the large mixing by 
using the hierarchical Dirac mass matrix in the see-saw mechanism.
This see-saw enhancement was studied previously by Smirnov and 
Tanimoto \cite{enhance}.        
\par    
For simplicity, we discuss the two family model.        
The maximal mixing is  generally given by the matrix    
\begin{equation}    
M_{LL}\simeq\left(\matrix{    
             a & b \cr    
             b & a \cr    
             }\right) \ ,    
\end{equation}    
where the eigenvalues are $a\pm b$.     
Consider two extreme cases.    
\noindent    
The first one  is the hierarchical case($a \simeq b$) which is given by the matrix     
\begin{equation}          
     M_{LL}^{hier}\sim\left (\matrix{    
          1 & 1 \cr 1 & 1\cr} \right ) \ .    
\end{equation}    
The second  one is the degenerate case($a \gg b$ or $a \ll b$)    
 which is generally given by the     
matrices     
\begin{equation}    
	M_{LL}^{deg1}\sim\left (\matrix{    
          1 & \e \cr \e & 1\cr} \right ) \quad {\rm or}        
	\ \  M_{LL}^{deg2}\sim\left (\matrix{    
          \e & 1 \cr 1 & \e\cr} \right ) \ ,        
\label{natural}    
\end{equation}        
where $\e\ll 1$.    
 Since the hierarchical case with large angle solution is already excluded     
from the naturalness in section 2,    
 we discuss  only the degenerate case in this section.    
The inverse matrix of $M_{LL}^{deg1}$ in eq.(\ref{natural})     
is     
\begin{equation}    
M_{LL}^{-1}\sim \left(\begin{array}{cc}    
 1 & -\e \\    
 -\e & 1 \\ \end{array}\right).    
\end{equation}    
If we consider the charged lepton or down quark mass matrix 
\begin{equation}  
M_{LR}  
\sim \left(\begin{array}{cc}    
 a\l^2 & b\l \\    
 b\l & 1 \\ \end{array}\right) \quad {\rm with}\ \ a,b\simeq{\cal O}(1) \ ,    
\label{MLD}    
\end{equation}    
\noindent    
as a typical neutrino Dirac mass matrix,     we obtain    
the right-handed Majorana mass matrix    
\begin{equation}    
M_{RR}=M_{LR}^TM_{LL}^{-1}M_{LR}\sim\left(\begin{array}{cc}    
 \l^2(b^2-2ab\e\l+a^2\l^2) & \l(b-(a+b^2)\e\l+ab\l^2) \\    
 \l(b-(a+b^2)\e\l+ab\l^2)  & 1-2b\e\l+b^2\l^2 \\ \end{array}\right) \ .    
 \label{majorana}    
\end{equation}    
\noindent    
Then the large angle solution like eq.(\ref{natural}) is obtained in the 
case when the factor $\e\l$ of all elements in $M_{RR}$ is  fine tuned     
to the leading term.    
Similar argument is valid also for the case of  $M_{LL}^{deg2}$.     
\par      
Since we must require the fine-tuning for the matrix $M_{RR}$, 
the solutions with the almost degenerate light neutrino masses,    
which are not excluded by the naturalness of the  mass matrix,      
could be excluded by the arguments of naturalness of right-handed Majorana mass matrix in the see-saw realization.
 \par 	        
In three family case, the situation is same as in the case of two families.    
Consider, for example, case(a-1) with the three degenerate masses.    
 Following eq.(\ref{MLL1}), we obtain         
\begin{equation}        
        M_{LL}^{-1}   \sim\left(\matrix{        
            1  & -\e & -\e \cr        
            -\e & 1 & -\e \cr        
            -\e & -\e & 1 \cr        
                          }\right)\ ,          
\end{equation}        
 \noindent where $\e_i$'s are taken to be equal to $\e$.    
If we accept  the following natural mass matrices in the quark sector \cite{Natu}         
\begin{equation}          
      M_u \propto \left (\matrix{\l^7 & \l^6 & \l^4\cr         
	  \l^6  & \l^4  & \l^4 \cr  \l^4 &  \l^4 & 1 \cr} \right ) \ ,        
	     \qquad         
	  M_d \propto \left (\matrix{\l^4 & \l^3 & \l^5\cr         
	  \l^3  & \l^2  & \l^2 \cr  \l^5 &  \l^2 & 1 \cr} \right ) \ ,        
\end{equation}        
 for the Dirac neutrino mass matrix, 
\noindent       
 we obtain        
\begin{equation}        
     M_{RR}\propto \left(\matrix{        
              \l^8(1+2\l^2\e) & \l^8(1+\e) & \l^4(1+\l^2\e) \cr        
              \l^8(1+\e) & 2\l^8(1+\e) & \l^4(1+\e) \cr        
              \l^4(1+\l^2\e) & \l^4(1+\e) & 1+4\l^4\e    \cr        
               }\right) \qquad {\rm for} \quad M_{LR}=M_u \ ,     
   \end{equation}        
and   
\begin{equation}        
     M_{RR}\propto \left(\matrix{        
              \l^6(1+2\l\e) & \l^5(1+\e) & 2\l^3(\l^2+\e) \cr        
              \l^5(1+\e) & 2\l^4(1+\e) & \l^2(1+\e) \cr        
              2\l^3(\l^2+\e) & \l^2(1+\e) & 1+2\l^2\e    \cr        
               }\right) \qquad {\rm for} \quad M_{LR}=M_d \ .     
   \end{equation}  
\noindent    
It is found that all the elements in $M_{RR}$   
for both cases are  fine-tuned  to the leading term.    
Thus, if we take the natural quark mass matrix for the neutrino Dirac 
mass matrix,        
  which may be justified in GUT's model, we will not be able to
 obtain the natural         
neutrino mass matrix for the case with the large angle mixing angle         
and the degenerate masses.        
\par        
$V_{2,11,12}$ 
with the hierarchical solution     
$m_3 \gg m_2 \simeq m_1$,  and $V_{3,6}$ with $m_3 \simeq m_2 \gg m_1$ 
are exactly the same situation as in the above degenerated case(a-1).     
The hierarchical realization with a large mixing angle needs precisely     
fine-tuned  right-handed Majorana mass structure.         
\par        
Let us consider another typical mixing pattern $V_9$ and $V_{10}$,         
where the large mixing angle  is excluded.        
Here the mass hierarchy is inverse.         
For $V_9$  the $\m$-like neutrino is the heaviest one, while the $\tau$-like 
neutrino is  the lightest one.           
For $V_{10}$ the $\m$-like neutrino is the heaviest one,  and the $e$-like 
neutrino is  the lightest one.          
This inverse mass hierarchy requires the huge mass hierarchy of        
 the right-handed Majorana masses.        
Assigning the numerical value to $V_9, V_{10}$ and         
neutrino masses with hierarchical structure $m_3 \gg m_2 \geq m_1$,        
we found that the huge right-handed Majorana mass hierarchy such as 
$M_1/M_3\simeq 10^4\sim 10^6$ may still be a reasonable hierarchy.        
 For example, in the case $V_9$ the left-handed Majorana masses are given by 
eq.(\ref{MLL9}) and eqs.(\ref{M9dg},\ref{M9hie}), from where we obtain        
\begin{equation}        
M_{LL}^{(9)} \propto\left(\matrix{1  &         \l^2 & \l^2 \cr        
                              \l^2 &  \l^4 & \l^4 \cr        
                              \l^2 & \l^4       & \l^4\cr} \right) \ ,\qquad
\left(\matrix{1  &         \l^4 & \l^2 \cr        
                              \l^4 &  \l^8 & \l^6 \cr        
                              \l^2 & \l^6       & \l^4\cr} \right) \ ,        
\end{equation}         
for $m_3 \gg m_2 \simeq m_1$ in eq.(\ref{M9dg}) and for 
$m_3 \gg m_2 \gg m_1$ in eq.(\ref{M9hie}), respectively.
Using this matrix    
and considering the hierarchical structure as a quark sector for the Dirac 
neutrino masses $M_{LR}$, we get  the right-handed Majorana mass matrices: 
\begin{eqnarray}        
M_{RR}^{(9)} &\propto & \left(\matrix{        
              \l^8 & \l^8 & \l^4 \cr        
              \l^8 & \l^8 & \l^4 \cr        
              \l^4 & \l^4 & 1 \cr}\right)     
                         \qquad {\rm for} \quad M_{LR}=M_u\ ,\nonumber\\        
         &\propto & \left(\matrix{        
              \l^6 & \l^5 & \l^3 \cr        
              \l^5 & \l^4 & \l^2 \cr        
              \l^3 & \l^2 & 1 \cr}\right)     
                         \qquad {\rm for} \quad M_{LR}=M_d     
\end{eqnarray}
for the case of $m_3 \gg m_2 \simeq m_1$ and 
\begin{eqnarray}        
M_{RR}^{(9)} &\propto & \left(\matrix{        
              \l^8 & \l^6 & \l^4 \cr        
              \l^6 & \l^4 & \l^2 \cr        
              \l^4 & \l^2 & 1 \cr}\right)     
                         \qquad {\rm for} \quad M_{LR}=M_u\ ,\nonumber\\        
         &\propto & \left(\matrix{        
              \l^6 & \l & \l \cr        
              \l & 1  & 1 \cr        
              \l & 1 & 1 \cr}\right)     
                         \qquad {\rm for} \quad M_{LR}=M_d     
\end{eqnarray}
for the case of $m_3 \gg m_2 \gg m_1$.        
For both cases we need the huge right-handed Majorana mass hierarchy like
$M_1/M_3\simeq {\cal O}(\l^{6\sim 8})\simeq {\cal O}(10^{4\sim 6})$.
We also obtain similar hierarchy for the case of $V_{10}$.   
Thus, the natural mass matrix is given by $V_9$ and $V_{10}$ on the second 
stage. 
\par       
We conclude that the atmospheric neutrino deficit cannot be reconciled 
with the naturalness.        
But what happens if the atmospheric neutrino deficit is indeed    
due to  the neutrino oscillation?        
This means that one should not take the quark mass matrix texture for the Dirac mass matrix     
of the neutrino.        
The GUT's relation is not valid at least for the neutrino mass matrix.        
But the flavor symmetry  such as $U(1)$ is favored in the framework of     
the see-saw mechanism. Actually,  quasi-degenerate neutrinos have been     
already obtained from an abelian group symmetry \cite{U1}.        
Zee mass matrix \cite{Zee} may be suggested as an alternative model without 
the see-saw mechanism.        
\section{Conclusion} 	        
We have studied the neutrino mass hierarchy and mixing in terms of        
naturalness of the mass matrix.        
We have succeeded to select a few neutrino mixing patterns,     
which could lead to the natural neutrino mass matrix.        
The mixing patterns $V_1$,     
$V_4$, $V_5$, $V_7$ and $V_8$ are unnatural for the solution with at least     
one hierarchical mass relation among three neutrinos.
On the contrary, the mixing patterns of $V_3, V_6, V_9, V_{10}, V_{11}$ and 
$V_{12}$ are the natural ones in the first stage of naturalness.
Also in the framework of the three degenerate mass pattern, mixing    
matrices $V_{1}$, $V_{5}$, $V_6$, $V_{11}$ and  $V_{12}$   
with the large mixing lead to the natural mass matrices     
in the first stage of naturalness.    
However, those patterns except for $V_9$ and $V_{10}$, lead to the unnatural 
right-handed Majorana mass matrix if the natural quark mass matrices     
are taken  for the Dirac neutrino mass matrix in the framework of     
the see-saw mechanism.        
\par	         
The mixing patterns $V_9$ and $V_{10}$ without the large mixing lead        
to the natural right-handed Majorana mass matrix.
In this case there should be a huge mass hierarchy such as 
${\cal O}(10^{4\sim 6})$ in order to reproduce the     
inverse mass hierarchy of the light neutrinos.        
\par	           
The naturalness of the neutrino mass matrix        
will be clarified in the near future due to the data from the numerous 
neutrino oscillation experiments being carried out.        
\vskip 0.5cm        
\noindent    
{\bf Acknowledgment}\\    
The authors would like to give special thanks to Prof. C. Jarlskog for 
her valuable comments and encouragements in writing this paper. 
We wish to thank Dr. T. Turova for the careful readings of the 
manuscript and important comments.     
Also we thank the kind hospitality at the Department of Mathematical Physics 
of LTH, Lund University, where this work was done.
One of the authors(M.M) expresses his thanks to the Japan Society of the Promotion of Sciences  and the Royal Swedish Academy of Sciences(RSAS) for the exchange program. Also this work is partially supported by a grant from the RSAS.     
        
\newpage
\begin{center}
\Large{\bf Table 1}
\end{center}
\begin{center}    
\begin{tabular}{|c|c|c|} \hline    
 case(a)	&Mass relation	& Naturalness	\\ \hline    
(1)$\quad V_1$& $m_3\gg m_2 \gg m_1$	& No	\\    
	& $m_3\gg m_2 \simeq m_1$	& No	\\    
	& $m_3 \simeq m_2 \simeq m_1$	& Yes	\\ \hline    
(2)$\quad V_2$& $m_3\gg m_2 \gg m_1$	& No	\\    
	& $m_3\gg m_2 \simeq m_1$	& Yes	\\    
	& $m_3 \simeq m_2 \simeq m_1$	& Yes(No:$2\b$ decay)	\\ \hline    
\end{tabular}\\    
\vspace{3mm}    
Table 1: The possible solutions from the naturalness for case(a).    
The exclusion from other experiments is shown in the parenthesis 
in the third  column.       
\end{center}
\begin{center}
\Large{\bf Table 2}
\end{center}       
\begin{center}    
\begin{tabular}{|c|c|c|} \hline    
Case(b)	& Mass relation	& Naturalness	\\     
\hline    
(1)$\quad V_3$	& $m_3\simeq  m_2\gg m_1$	& Yes  \\    
	            & $m_3\gg m_2\simeq m_1$	& No 	\\     
				& $m_3\gg  m_2\gg m_1$	   & No    \\  \hline    
(1)$\quad V_4$  & $m_3\gg  m_2\gg m_1$	   & No    \\    
	            & $m_3\gg  m_2\simeq m_1$	&No	   \\ \hline    
(2)$\quad V_5$	& $m_3\gg  m_2\gg m_1$	    &  No		\\     
                & $m_3\gg  m_2\simeq m_1$	 & No	  \\    
		        & $m_3\simeq  m_2\simeq m_1$	&Yes \\ \hline    
(2)$\quad V_6$	& $m_3\simeq  m_2\gg m_1$	& Yes		\\     
	            & $m_3\gg  m_2\simeq m_1$	&No     \\    
	            & $m_3\gg  m_2\gg m_1$	    &No     \\    
	         	& $m_3\simeq  m_2\simeq m_1$	&Yes	\\ \hline    
(3)$\quad V_7$	& $m_3\simeq  m_2\gg m_1$	& No	\\    
	            & $m_3\gg  m_2\simeq m_1$	&No	\\ \hline    
(3)$\quad V_8$	& $m_3\gg  m_2\gg m_1$ & No	\\    
	            & $m_3\gg  m_2\simeq m_1$	&No		\\ \hline    
(3)$\quad V_9$	& $m_3\gg  m_2\gg m_1$	    &   Yes	\\    
	            & $m_3\gg  m_2\simeq m_1$	&Yes	\\ \hline    
(3)$\quad V_{10}$&$m_3\gg  m_2\gg m_1$	& Yes	\\    
	            & $m_3\gg  m_2\simeq m_1$	&Yes	\\ \hline  
(3)$\quad V_{11}$	& $m_3\gg  m_2\gg m_1$	    &   No	\\    
	            & $m_3\gg  m_2\simeq m_1$	&Yes	\\ \hline    
(3)$\quad V_{12}$&$m_3\gg  m_2\gg m_1$	& No	\\    
	            & $m_3\gg  m_2\simeq m_1$	&Yes	\\ \hline    
\end{tabular}\\    
Table 2: The possible solutions from the naturalness for case(b).    
\end{center} 
\end{document}